\documentclass[prl,aps,showpacs,twocolumn,unsortedaddress]{revtex4}
\usepackage{graphics,bm}
\usepackage{amssymb}
\usepackage{epsfig}
\usepackage{epsf}

\begin{document}

\title{Quantitative Probe of Pairing Correlations in a Cold Fermionic Atom Gas}

\author{R.A. Duine}
\email{duine@physics.utexas.edu}
\homepage{http://www.ph.utexas.edu/~duine}

\author{A.H. MacDonald}
\email{macd@physics.utexas.edu}
\homepage{http://www.ph.utexas.edu/~macdgrp}

\affiliation{The University of Texas at Austin, Department of
Physics, 1 University Station C1600, Austin, TX 78712-0264}

\date{\today}

\begin{abstract} A quantitative
measure of the pairing correlations present
in a cold gas of fermionic atoms can be obtained by
studying the dependence of RF spectra on hyperfine state populations.
This proposal follows from a sum rule that relates the
total interaction energy of the gas to RF spectrum line
positions.  We argue that this indicator of pairing
correlations provides information comparable to that
available from the spin-susceptibility and NMR measurements common
in condensed-matter systems.
\end{abstract}


\maketitle

\def\bx{{\bf x}}
\def\bk{{\bf k}}
\def\half{\frac{1}{2}}
\def\args{(\bx,t)}

\noindent {\it Introduction} --- The realization of degenerate
atomic Fermi gases
\cite{demarco2001,truscott2001,schreck2001,granade2002,jochim2002,roati2002,hadzibabic2003}
has opened new opportunities for experimental discovery.  The focus to date has mainly been
on efforts to observe the condensation of
atomic Cooper pairs to form a superfluid state similar to the
BCS state of electrons in a superconductor \cite{stoof1996}.
Strategies for achieving observable pairing effects have so far
hinged on the occurrence of strong attractive atom-atom
interactions near a Feshbach resonance
\cite{stwalley1976,tiesinga1993,heinzen}, in which a molecular
bound state of one atom-atom scattering channel is close to
the two-atom continuum threshold of another.  The proximity of a Feshbach
resonance can be adjusted by tuning a magnetic bias field,
drastically altering the scattering behavior of atoms and
allowing the $s$-wave scattering length to be varied over values
corresponding to effective interactions that are weak or strong,
and repulsive or attractive. The scattering length, henceforth
denoted by $a$\,, completely characterizes the interaction
properties of atoms at low temperatures and low densities. The
Feshbach resonance makes it possible to study one of the paradigms
\cite{leggett1980,nozieres1985,footnote} of fermion pairing
theory, the BCS-BEC crossover, experimentally.

Experimental groups have already observed the formation of thermal
gases \cite{strecker2003,cubizolles2003} and Bose-Einstein
condensates \cite{jochim2003,greiner2003,zwierlein2003} of
diatomic molecules. Condensation of fermionic atom pairs on the
attractive interaction side of the resonance, where there is no
two-atom bound state and the analogy to the BCS-BEC crossover
problem is closer, has also been reported
\cite{regal2004,zwierlein2004}. Because the BCS transition does
not manifest itself strongly \cite{houbiers1997} in the expanded
density profile of the gas there is a need for quantitative and
direct measurements of pairing correlations, one that has
motivated a large number of proposals. Several works focused on
the change of light scattering due to the transition from the
normal to the superfluid phase as a detection method
\cite{zhang1999,ruostekoski1999,weig2000,bruun2004}.  Later,
resonant laser light was proposed to induce tunnelling between the
superfluid and normal state of the gas \cite{torma2000}. The
experimental realization of Bragg spectroscopy in a Bose gas
\cite{stamperkurn1999} has inspired theoretical work on pairing
effects in the dynamic structure function
\cite{minguzzi2001,buchler2004}.  Other interesting proposals
include ones based on pairing induced changes in collective-mode
frequencies
\cite{baranov2000,zambelli2001,bruun2001,regal2004b,kinast2004,bartenstein2004},
rotational properties of the gas \cite{farine2000,cozzini2003},
the expansion of the gas \cite{menotti2002}, and atomic density
noise-correlation properties \cite{altman2004}.

In this Letter we propose a more direct probe of pairing
correlations that is similar to the spin magnetic susceptibility
and nuclear-spin relaxation probes commonly used to detect
electron pairing in condensed matter systems, and is able to
detect pairing even when it does not lead to long-range coherence.
We suggest a measurement of the cost in interaction energy when
the number of Cooper pairs in the system is reduced by making the
hyperfine state populations unequal.  As we show below, this
energy change can be extracted from data obtained using the RF
spectroscopy techniques that have already been developed by
several experimental groups \cite{chin2004,gupta2003,regal2003}.
To illustrate the direct relationship between the hyperfine
population dependence of the interaction energy and pairing
correlations, we compare the predictions of BCS theory for this
quantity with its predictions for the more familiar magnetic
susceptibility probe, which measures instead the dependence of
total (interaction plus kinetic) energy on the same variable.

\noindent {\it Interaction Energy Sum Rule} --- We consider a gas
of fermionic atoms that consists of a mixture of two hyperfine
states denoted  by $|\!\uparrow\rangle$ and
$|\!\downarrow\rangle$. We assume the temperatures to be low
enough so that only $s$-channel interactions, forbidden between
atoms in the same hyperfine state by the Pauli principle, are
significant.  The Hamiltonian of the gas is therefore
$H=H_0+H_{\rm int}$, where the noninteracting part is
\begin{equation}
 H_0 = \int\!d\bx \sum_{\alpha=\left\{ \uparrow,\downarrow \right\}}
\psi^\dagger_\alpha (\bx)\left(- \frac{\hbar^2 {\bf
\nabla}^2}{2m}+\epsilon_\alpha \right) \psi_\alpha (\bx)~,
\end{equation}
and $\psi_\alpha$ is the fermionic annihilation operator for
hyperfine state $|\alpha\rangle$. The internal Zeeman energy of a
hyperfine state is denoted by $\epsilon_\alpha$, and for
simplicity we have neglected any inhomogeneity of the magnetic
field. In particular, this implies that we neglect the effects of
the magnetic trapping potential. We take the interaction between
unlike hyperfine states to be a contact interaction with strength
$V_{\uparrow\downarrow}$, which should be chosen to produce the
correct two-body scattering amplitude \cite{footnote2}. With these
assumptions, the interaction part of the Hamiltonian is
\begin{equation}
 H_{\rm int} = V_{\uparrow\downarrow} \int\!d\bx\psi^\dagger_\uparrow (\bx)
\psi^\dagger_\downarrow (\bx)
 \psi_\downarrow (\bx) \psi_\uparrow(\bx)~.
\end{equation}

In the RF experiments one of the system hyperfine species (say $|\!\downarrow\rangle$) is
coupled to a spectator hyperfine state ($|{\rm
s}\rangle$), and the number of atoms in the spectator state, $N_{s}$, is
detected as a function of the frequency of the RF-field.  In the linear
response limit $N_s$ is proportional to the rate of  $|\!\downarrow\rangle \to |{\rm
s}\rangle$, transitions which we denote by $I(\omega)$.  We define the position of the associated
RF spectrum absorption line as
\begin{equation}
\label{lineposition}
\hbar \bar \omega \equiv \frac{\int\! d \omega \hbar \omega
I(\omega)}{\int d\omega I(\omega)}~.
\end{equation}
Using a formal golden-rule expression, $I(\omega)$ can be
expressed in terms of a two-particle correlation function of
Fermion fields.  It then follows from the fermion analog of sum
rules derived in Refs.~\cite{pethick2001,oktel2002} that
\begin{equation}
\label{eq:shift}
 \hbar \bar \omega = \hbar \omega_0
 + \frac{1}{n_\uparrow} \left( V_{\downarrow\uparrow} - V_{{\rm s}\uparrow}
 \right) \langle \psi^\dagger_\uparrow \psi^\dagger_\downarrow
\psi_\uparrow \psi_\downarrow \rangle~,
\end{equation}
where $V_{{\rm s}\uparrow}$ denotes the strength of the contact
interaction between the spectator and $|\!\uparrow\rangle$
hyperfine states, and $n_\alpha$ denotes the average density in
spin state $|\alpha\rangle$. The shift of $\hbar \bar \omega$ from
the bare line position $\hbar
\omega_0=\epsilon_\downarrow-\epsilon_{\rm s}$ differs from the
interaction energy per volume by a factor,
$V_{\downarrow\uparrow}n_{\uparrow}/(
V_{\downarrow\uparrow}-V_{{\rm s}\uparrow})$, which can be held
constant through the experiments and if necessary can be
accurately determined by separate measurements. We conclude that
the RF spectra enable a direct measurement of the interaction
energy density
\begin{equation}
\label{eq:eint}
  e_{\rm int} (n_\uparrow,n_\downarrow) \equiv
  V_{\uparrow_\downarrow} \langle \psi^\dagger_\uparrow \psi^\dagger_\downarrow
\psi_\downarrow \psi_\uparrow \rangle~.
\end{equation}

\noindent
{\it Pairing and Interaction Energy in BCS Theory} --- The inverse spin-susceptibility of an
unpolarized system of spin-$1/2$ particles may be expressed in terms of the dependence
of free energy on spin-polarization:
\begin{equation}
\label{eq:spinsusc}
  \chi_{\rm s}^{-1} = \left . \frac{\partial^2 f_{\rm tot}}{\partial \delta n^2}\right|_{n}
  =  \frac{1}{2} \left[ \frac{\partial^2 f_{\rm tot}
  \left( n_\uparrow,n_\downarrow \right)}{\partial n_\uparrow^2}
 -  \frac{\partial^2 f_{\rm tot}
  \left( n_\uparrow,n_\downarrow \right)}{\partial n_\uparrow \partial
  n_\downarrow}\right],
\end{equation}
where $f_{\rm tot}$ is the total free energy per unit volume of
the gas, $n$ is the total density, and $\delta n \equiv
n_{\uparrow}-n_{\downarrow}$ is the spin density. It is well-known
that the $\chi_{\rm s}$ is strongly suppressed when atoms can gain
energy by pairing.
 ($\chi_{\rm s}$ vanishes as $T \to
0$ in the BCS state.)  For attractive atom-atom interactions, the
energy cost of finite-spin polarization has positive contributions
from both interaction and kinetic energy. In the following we
compare the spin-susceptibility with an alternate quantity that is
defined in terms of the interaction energy alone and can be
extracted from RF spectroscopy experiments performed for a series
of hyperfine-state populations:
\begin{equation}
\label{eq:spinsuscint}
 \chi^{-1}_{\rm{int,s}} =\left. \frac{\partial^2 e_{\rm int} (n_\uparrow,n_\downarrow)}{\partial \delta
  n^2}\right|_n ~.
\end{equation}
As we show below, this quantity and the inverse
spin-susceptibility provide similar probes of pairing
correlations.

We evaluate this {\em interaction susceptibility} using BCS theory from
which it follows that
\begin{equation}
\label{eq:eintbcs}
   e_{\rm int} (n_\uparrow,n_\downarrow)= \frac{|\Delta|^2}{V_{\uparrow\downarrow}}
    + \frac{4\pi a \hbar^2 n_\uparrow
n_\downarrow}{m}~,
\end{equation}
where the dependence of the gap $\Delta \equiv
V_{\uparrow\downarrow} \langle \psi_\downarrow \psi_\uparrow
\rangle$ on temperature and hyperfine densities can be determined
by solving the self-consistent mean-field equations. The
mean-field Hamiltonian is \cite{houbiers1997}
\begin{widetext}
\begin{eqnarray}
\label{eq:mfham}
 H_0 &=& \int\!d\bx \left\{ \psi^\dagger_\uparrow (\bx)\left(- \frac{\hbar^2 {\bf
\nabla}^2}{2m}+\frac{4\pi a \hbar^2 n_\downarrow}{m}-\mu_\uparrow
\right) \psi_\uparrow (\bx) +
 \psi^\dagger_\downarrow (\bx) \left(
  - \frac{\hbar^2 {\bf
\nabla}^2}{2m} + \frac{4\pi a \hbar^2 n_\uparrow}{m}
-\mu_\downarrow
\right)\psi_\downarrow (\bx) \right. \nonumber \\
&+& \left. \Delta \psi^\dagger_\uparrow (\bx)
\psi^\dagger_\downarrow (\bx)
 +\Delta^* \psi_\downarrow (\bx) \psi_\uparrow(\bx)-\frac{\left|
\Delta\right|^2}{V_{\uparrow\downarrow}} -\frac{4\pi a \hbar^2
n_\uparrow n_\downarrow}{m} \right\}~,
\end{eqnarray}
\end{widetext}
Note that the renormalization $V_{\uparrow\downarrow} \to 4 \pi a
\hbar^2/m$ can be made at this stage in the Hartree mean-field potential.
The chemical potentials of the two hyperfine
states are denoted by $\mu_\alpha$, and are not necessarily equal,
thus allowing for a density difference between the two hyperfine
states.

The partial densities are given by
\begin{equation}
\label{eq:densities}
  n_{\alpha} = \int\!\frac{d\bk}{(2\pi)^3} \left\{|u_\bk|^2 N(\hbar \omega_{\bk,\alpha})+
  |v_\bk|^2 \left[ 1- N(\hbar \omega_{\bk,-\alpha})\right]
\right\}~,
\end{equation}
where $u_\bk$ and $v_\bk$ are the Bogoliubov coherence factors,
$N(x)=\left[e^{\beta x}+1\right]^{-1}$ is the Fermi distribution
function, and the $|\uparrow\rangle$ quasiparticle dispersion is given by
\begin{equation}
 \hbar \omega_{\bk,\uparrow} = \frac{\mu'_\downarrow-\mu'_\uparrow}{2}
  + \sqrt{[\epsilon_\bk-(\mu'_\uparrow+\mu'_\downarrow)/2]^2+|\Delta|^2}~.
\end{equation}
An identical expression, with the hyperfine labels
interchanged, applies for  $\hbar \omega_{\bk,\downarrow}$. The
Hartree-Fock mean-field shift is absorbed in the chemical
potential via  $\mu'_\alpha = \mu_\alpha - 4\pi a \hbar^2
n_{-\alpha}/m$.  These equations for the densities need to be solved together with
the BCS gap equation
\begin{equation}
\label{eq:gapeq}
 \int \frac{d\bk}{(2\pi)^3}
\frac{1-N(\hbar\omega_{\bk,\uparrow})-N(\hbar\omega_{\bk,\downarrow})}{2\sqrt{[\epsilon_\bk
-(\mu_\uparrow+\mu_\downarrow)/2]^2+|\Delta|^2}}
 =-\frac{1}{V_{\uparrow\downarrow}}~.
\end{equation}
Eq.~(\ref{eq:gapeq}) contains a ultraviolet divergence that is
renormalized by using introducing the $s$-wave scattering length
($a$) between $|\uparrow\rangle$ and $|\downarrow\rangle$
hyperfine states by means of the Lippmann-Schwinger equation
\begin{equation}
\label{eq:lseq}
 \frac{m}{4\pi a \hbar^2} = \frac{1}{V_{\uparrow\downarrow}}
  +  \int_{\bk \leq k_\Lambda} \frac{d \bk}{(2\pi)^3} \frac{1}{2\epsilon_\bk}~,
\end{equation}
where we have introduced an ultraviolet cutoff $k_{\Lambda}$. It
follows \cite{houbiers1997} from Eq.~(\ref{eq:lseq}) that
\begin{equation}
 V_{\uparrow\downarrow} = \frac{4\pi a
\hbar^2}{m} \frac{1}{1-\frac{2a k_{\Lambda}}{\pi}}~.
\end{equation}
For $k_\Lambda \simeq (100 a_0)^{-1}$, and $a=-2000 a_0$, we find
that $V_{\uparrow\downarrow} \simeq 0.07 \times (4\pi a
\hbar^2/m$).  Note that although the spin densities and the BCS
gap parameter are independent of the short-range properties of the
interatomic potential, the interaction energy in
Eq.~(\ref{eq:eintbcs}) is not \cite{houbiers1997}.

\begin{figure}
\includegraphics{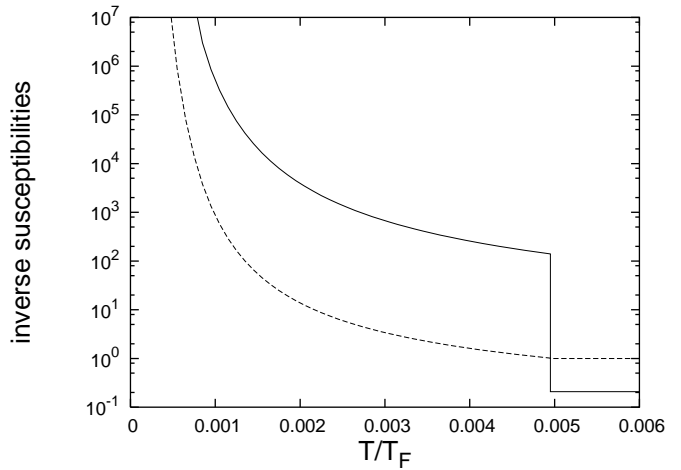}
\caption{\label{fig:ddeint} Plot of $\chi_{\rm int,s}^{-1}$ (solid
line) and $\chi_{\rm s}^{-1}$ (dashed line) {\em vs.} temperature.
Both susceptibilities are plotted in units of the quasiparticle
density-of-states  $(\partial n/\partial \mu)$. Note that the
mean-field-theory interaction susceptibility is discontinuous at
the critical temperature.}
\end{figure}

Above the critical temperature, $T_{\rm BCS} \simeq 0.6 T_{\rm F} \;
e^{-\pi/2 k_{\rm F}|a|}$, $\Delta \to 0$ so that
\begin{equation}
 e_{\rm int} (n_\uparrow,n_\downarrow) =
    \frac{4\pi a \hbar^2 n_\uparrow
n_\downarrow}{m}~,
\end{equation}
and $\chi_{\rm int,s}^{-1} = 2 \pi |a| \hbar^2 /m$ is temperature
independent. For $T < T_{\rm BCS}$ the interaction energy is given
by Eq.~(\ref{eq:eintbcs}). Linearization of the gap equation
implies that for $T \uparrow T_{\rm BCS}$,
\begin{equation}
\label{eq:eintbcsagain}
 \chi_{\rm int,s}^{-1}  = \frac{\pi}{0.07 k_{\rm F}
|a|} \left( \frac{\partial n}{\partial \mu}\right)^{-1} + 2 \pi
|a| \hbar^2 /m.
\end{equation}
In Fig.~\ref{fig:ddeint} we plot $\chi_{\rm int,s}^{-1}$ and the
inverse susceptibility {\em vs.} temperature, for the case of
$n_\uparrow=n_\downarrow=n/2$, scattering length $a=-2000a_0$
where $a_0$ is the Bohr radius, and density $n=10^{12}$ cm$^{-3}$.
For these parameters $T_{\rm BCS} \simeq 0.005 \times T_{\rm F}$.
The inverse interaction susceptibility is greatly enhanced for
temperatures below the BCS transition temperature, thus providing
a clear signature of atomic Cooper pairs.

{\it Discussion and Conclusions} --- Although we have so far assumed
a homogeneous Fermi gas, we believe that our results
also apply to the case of a trapped, and therefore, inhomogeneous
Fermi gas. There are two main reasons for this. First, since the
inhomogeneity in the density profile leads to broadening of the RF
spectrum, the sum rule in Eq.~(\ref{lineposition}) is valid with
the density $n_\uparrow$ replaced by a mean density
\cite{gupta2003,regal2003}. Second, because the inverse Fermi wave
vector is much smaller than the harmonic oscillator length for the
experimental systems of interest, the system may be treated within
a local density approximation \cite{houbiers1997}. Since the BCS
gap parameter $\Delta$ will be largest in the center of the trap,
the result in Eq.~(\ref{eq:eintbcsagain}) should be evaluated at
the center of the trap. With these two modifications, the results
presented in this Letter should carry over, in a more than
qualitative manner, to the inhomogeneous case.

For strong attractive interactions, $k_F |a| \sim 1$, mean-field
theory is not expected to be accurate.  In particular, the
superfluid transition temperature is expected to be limited by the
loss of long range coherence rather than by the thermodynamics of
pair formation.  The thermodynamic probe we discuss here is
sensitive to the occurrence of pair correlations and not
particularly sensitive to the establishment of long range
coherence.  It should therefore be able to detect the gradual
development of pairing correlations with increasing interaction
strength as the superfluid state is approached. We note that
Bourdel {\it et al.} \cite{bourdel2003} have measured the ratio of
the interaction energy and kinetic energy of a Fermi gas by
comparing expanded density profiles of an interacting gas of atoms
with expansion profiles of a gas at zero scattering length. In the
weak-coupling limit such a measurement would provide direct
information on the temperature dependence of the interaction
energy, since the kinetic energy is almost independent of
temperature in this case. In the strong-coupling limit it is,
however, not clear how the kinetic energy depends on the density
difference, and it is, therefore, not obvious that a measurement
of the ratio of interaction and kinetic energy for different
hyperfine state populations would provide a sensitive probe of
pairing correlations in the gas in this limit. Finally we remark
on the possibility of realizing inhomogeneous pair-condensate
states \cite{FFLO} in cold atom systems with unbalanced hyperfine
state populations.  These states could be detected by bringing the
system to equilibrium in a rotating reference frame and
visualizing their unusual \cite{FFLOVL} vortex-lattice structures.

This work was supported by the National Science Foundation under grant
DMR-0115947 and by the Welch Foundation.  We are grateful for informative
conversations with D.S. Jin.


\begin{thebibliography}{99}
\bibitem{footnote2} Since we do not include a molecular
field, that is coupled to the atomic field, in the description of
the interaction properties of the gas, we do not explicitly
consider a Feshbach resonance. However, our results do apply to
the case of a Feshbach resonance, provided we are far enough from
the resonance.
\end{thebibliography}
\end{document}